\documentclass[a4paper,twoside,10pt]{letter}
\usepackage{graphicx,saj,multicol,subeqnarray}

\def\papertype{Editorial}

\begin{document}
\baselineskip=3.1truemm
\columnsep=.5truecm
\newenvironment{lefteqnarray}{\arraycolsep=0pt\begin{eqnarray}}
{\end{eqnarray}\protect\aftergroup\ignorespaces}
\newenvironment{lefteqnarray*}{\arraycolsep=0pt\begin{eqnarray*}}
{\end{eqnarray*}\protect\aftergroup\ignorespaces}
\newenvironment{leftsubeqnarray}{\arraycolsep=0pt\begin{subeqnarray}}
{\end{subeqnarray}\protect\aftergroup\ignorespaces}
%


\markboth{\eightrm ON THE RADIO SPECTRUM OF SNR W44}{\eightrm D. Oni\'c}

{\ }



{\ }


\title{ON THE INTEGRATED CONTINUUM RADIO SPECTRUM OF SUPERNOVA REMNANT W44 
(G34.7-0.4): NEW INSIGHTS FROM PLANCK}


\authors{D. Oni\'c}

\vskip3mm


\address{Department of Astronomy, Faculty of Mathematics,
University of Belgrade\break Studentski trg 16, 11000 Belgrade,
Serbia}

\Email{donic}{math.rs}




\summary{In this paper, the integrated continuum radio spectrum of supernova remnant (SNR) W44 was analyzed up to 70 GHz, 
testing the different emission models that can be responsible for its particular shape. The observations by the 
{\footnotesize \textit{Planck}} space telescope made possible to analyze the high frequency part of radio emission from 
SNRs. Although the quality of radio continuum spectrum (a high scatter of data points at same frequencies) prevents 
us to make definite conclusions, the possibility of spinning dust emission detection towards this remnant is emphasized. 
In addition, a concave-down feature, due to synchrotron losses, can not be definitely dismissed by the present knowledge 
of the integrated radio continuum spectrum of this SNR.}


\keywords{ISM: individual (W44) -- ISM: supernova remnants -- radio continuum: ISM}

\begin{multicols}{2}
{


\section{1. INTRODUCTION}

Supernova remnant W44 (G34.7-0.4, 3C392) represents one of the most interesting SNRs. It is a member of the 
mixed-morphology (thermal composite) class of SNRs, characterized by a bright non-thermal shell-like radio 
morphology and centrally concentrated thermal X-ray emission (Rho \& Petre 1998; Vink 2012). W44 is a middle-aged SNR 
(possibly around $20\ 000$ yrs old), that has an asymmetric, quasi-elliptical shell morphology, presumably due to expansion 
in an inhomogeneous interstellar medium (Cardillo et al.\@ 2014). It is about half a degree in size ($35'\times27'$) 
at a distance of around 3 kpc, and it is located in a complex region of the inner Galactic plane rich in both thermal and 
non-thermal sources (Castelletti et al.\@ 2007). In fact, this remnant is well placed in the W48 molecular cloud complex, 
a rich star-forming region.

SNR W44 constitutes one of the few demonstrated cases of an SNR-molecular cloud interacting system (Castelletti et al.\@ 2007, 
and references therein). It is also detected in $\gamma$-rays. Cardillo et al.\@ (2014) concluded that 
$\gamma$-ray emission is most probably caused by the neutral pion decay (hadronic scenario). They also determined that the 
average gas density of the regions emitting 100 MeV - 10 GeV $\gamma$-rays is relatively high ($250-300\ \mathrm{cm}^{-3}$). 
Yoshiike et al.\@ (2013) confirmed that the W44 SNR environment consists of both molecular and atomic hydrogen and 
they concluded that molecular clouds, which are likely associated with this remnant, surround the radio shell. Their 
analysis also supports a hadronic origin of the $\gamma$-rays.

The global integrated radio continuum spectral index of this SNR is $\alpha = 0.37$ (Green 2014), where spectral 
index is defined as $S_{\nu}\propto\nu^{-\alpha}$, and $S_{\nu}$ is the flux density. This is significantly less than 0.5, 
the value that is predicted by the test-particle diffusive shock acceleration theory (DSA, Uro\v sevi\'c 2014, and 
references therein). On the other hand, such values for spectral index are common for majority of mixed-morphology 
SNRs (Oni\'c 2013, and references therein). Uchida et al.\@ (2012) reported another common feature for mixed-morphology 
SNRs: radiative recombination continua of highly ionized atoms in X-ray spectrum. They concluded that the spectrum is 
well reproduced by a thermal plasma in a recombining phase.

The spatially resolved radio spectral index study revealed that the detected localized absorption has a negligible 
influence on the total integrated flux, and thus has no measurable impact on the integrated continuum spectrum 
(Castelletti et al.\@ 2007). The most likely explanation of the particular local spectral inversion is the 
low-frequency radio continuum free-free absorption from ionized gas in the post shock region at the 
SNR/molecular cloud interface. An alternative explanation is that the thermal absorption is occurring inside 
the boundaries of the coincident \mbox{H\,{\sc ii}} region, along its periphery where the thermal electron 
density might be the highest. The spectral inversion is probably produced by a combination of both of these effects 
(Castelletti et al.\@ 2007).

It is worth mentioning that there is no evidence in the radio continuum spectrum of any coupling between the associated 
pulsar PSR B1853+01 and the surrounding SNR shell that could, for example, be observed as a gradual steepening from the 
pulsar to the shell (Castelletti et al.\@ 2007). In addition, the pulsar wind powers a small synchrotron nebula observed 
at radio frequencies and X-rays (Anderl et al.\@ 2014, and references therein).

A very interesting discovery of hard X-ray emission from SNR, not connected with associated pulsar or correspondent pulsar 
wind nebula, was reported by Uchida et al.\@ (2012). They found that hard X-rays have an arc-like structure 
spatially-correlated with a radio continuum filament. Uchida et al.\@ (2012) noted that the surface brightness distribution 
shows a clear anti-correlation with $^{12}$CO($J=2-1$) emission from a molecular cloud. Finally, they concluded that 
the hard X-rays are most likely due to a synchrotron enhancement in the vicinity of the cloud. In fact, the localized 
non-thermal X-ray emission, as it is seen in the case of W44, probably reflects the physics of shock propagation in 
a clumpy medium. In an inter-clump gas with a lower density, cloud shocks can propagate much faster than in the dense 
clumps, so they can possibly accelerate high-energy electrons capable enough to emit X-rays (Lee et al.\@ 2015).

Using {\it Spitzer} mid-IR observations, an \mbox{H\,{\sc ii}} region, located just outside the southeastern limb of SNR W44, 
was detected by Castelletti et al.\@ (2007). They showed that the combination of $8$ and $24\ \mu\mathrm{ m}$ {\it Spitzer} images 
reveals the hot dust grains in the Str\"{o}mgren sphere, limited to the east by an annular photo dissociation region (PDR) 
dominated by polycyclic aromatic hydrocarbons (PAHs) emitting near $8\ \mu$m. Castelletti et al.\@ (2007) noted that exactly 
at the interface between observed \mbox{H\,{\sc ii}} region and W44, as seen in the plane of the sky, a young stellar object 
is present. In fact, two massive young stellar objects are identified at the border of the \mbox{H\,{\sc ii}} region that 
is evolving within a molecular cloud shocked by the SNR W44 (Paron et al.\@ 2009; Ortega et al.\@ 2010).

Recently, the observations from a microwave survey of Galactic SNRs made by {\it Planck}\footnote{\textit{Planck} is a project 
of the European Space Agency (ESA) with instruments provided by two scientific consortia funded by ESA member states, 
with contributions from NASA (USA) and telescope reflectors provided by a collaboration between ESA and a scientific 
consortium led and funded by Denmark.} were published (Planck Collaboration Int.\@ XXXI 2014). {\it Planck} observed the sky in nine 
frequency bands covering 30 - 857 GHz with high sensitivity and a range of angular resolutions from $31'$ to $5'$ (Planck Collaboration 
Int.\@ XXXI 2014 and references therein). The Low Frequency Instrument covers the 30, 44, and 70 GHz bands while the 
High Frequency Instrument covers the 100, 143, 217, 353, 545, and 857 GHz.

Planck Collaboration Int.\@ XXXI (2014) reported that the synchrotron emission from SNR W44 is detected at levels 
above the emission from nearby unrelated regions at 30-70 GHz. They also noted that the measured 70 GHz flux density 
from \textit{Planck} is somewhat lower than that expected from the radio power law, while the 30 GHz flux density is higher. Planck Collaboration Int.\@ XXXI (2014), on the 
other hand, discounted the 30 GHz flux density due to possible confusion with unrelated large-scale emission from the 
Galactic plane. Also, it must be noted that the {\it Planck's} flux density measurements above 100 GHz are contaminated 
by an unrelated foreground emission. Planck Collaboration Int.\@ XXXI (2014) identified a particular structure at 
the eastern border of the radio SNR as a compact \mbox{H\,{\sc ii}} region, unrelated to the W44 SNR, but possibly a 
member of the same OB association as the progenitor.

In this paper, we have analyzed the radio spectrum of the W44 SNR up to 70 GHz, testing different emission models 
in order to replicate its particular spectral shape.

\section{2. ANALYSIS AND RESULTS}

Flux densities at different frequencies for SNR W44 were taken from Table 2 of Castelletti et al.\@ (2007), Table 1 of 
Sun et al.\@ (2011) and Table 3 of Planck Collaboration Int.\@ XXXI (2014) in the frequency range from 610 MHz to 70 GHz. 
Only flux densities with errors $<20\%$ were used in the analysis and only data corrected to the scale of 
Baars et al.\@ (1977) are taken from Castelletti et al.\@ (2007). Finally, the misprint in Table 2 of Castelletti 
et al.\@ (2007) regarding the data uncertainties taken from Altenhoff et al.\@ (1970) is taken into account - uncertainties 
are set to 10\% of the selected flux densities.

The significant scatter of flux densities measured at the same frequencies makes the integrated continuum radio spectrum difficult to 
analyze (see Figure 6 from Castelletti et al.\@ 2007 and Figures in this paper). The data come from a wide variety of 
telescopes with different beam sizes, which can have a significant impact on flux density measurements, since W44 lies within 
a particularly complex region close to the Galactic plane. The effects of possible contamination by the unrelated sources can be 
significant enough to account for the scatter in the radio continuum spectrum of W44.

\subsection{2.1. The synchrotron emission from the SNR W44}

The usual interpretation of the radio continuum spectrum of SNRs is that of a simple power law that arises from the 
synchrotron emission of charges accelerated by a diffusive shock acceleration (DSA) mechanism (Uro\v sevi\'c 2014, and 
references therein). In that case, the flux density is given by the following relation (Eq.\@ 1) \begin{equation}
S_{[\mathrm{Jy}]}(\nu) = S_{[\mathrm{Jy}]}(1\ \!\!\mathrm{GHz})\ \nu_{[\mathrm{GHz}]}^{-\alpha},                     
\end{equation}
where $\alpha$ is the radio spectral index. On the other hand, the high frequency part of the W44 radio spectrum clearly 
indicates that this simple model does not adequately explain the integrated continuum of this SNR. The lower value 
of the flux density at 70 GHz could be a possible indication of high frequency spectral bending (Planck Collaboration 
Int.\@ XXXI 2014) that can be represented by (Eq.\@ 2) \begin{equation}
S_{[\mathrm{Jy}]}(\nu) = S_{[\mathrm{ Jy}]}(1\ \!\!\mathrm{GHz})\ \nu_{[\mathrm{GHz}]}^{-\alpha}\ e^{-\frac{\nu}{\nu_{0}}},                     
\end{equation}
where $\nu_{0}$ is a characteristic cut-off frequency.

We performed least-squares fits to the radio spectrum (Fig.\@ 1) using a simple synchrotron model (solid line) as well 
as a synchrotron model with exponential cutoff (dotted line). The thick and thin lines in Figure 1 correspond to fits 
without and with the inclusion of the \textit{Planck} 30 GHz flux density, respectively. \textit{Planck} data is 
represented with a diamond symbol, except for the data point at 30 GHz, which is labeled with a triangle. The best 
fitting parameters for these two models are presented in Table 1. The asterisk in Table 1 indicates the cases when 
{\it Planck} data at 30 GHz is included in the fit.

It is known that the radio spectra of evolved SNRs could appear in a concave-down form. This kind 
of spectrum can be explained using DSA theory with the effect of synchrotron losses within the finite 
emission region. Generally, the concave-down form of the radio spectra should correspond to very old SNRs, 
for which electrons have had enough time to lose a significant amount of energy at the highest radio 
frequencies, and primarily to distant (mainly extragalactic) SNRs for which the limitation in telescope 
resolution leads to the capturing of radio emission from the sample of "exhausted" electrons (Uro\v sevi\'c 
2014, and references therein). The angular resolution of {\it Planck's} Low Frequency Instrument (LFI) at 
30, 44 and 70 GHz is $33'$, $24'$, $14'$, respectively (Mandolesi et al.\@ 2010). For a comparison, the angular 
resolution of Wilkinson Microwave Anisotropy Probe (WMAP) at 23, 33, 41, 61 and 94 GHz is around $53'$, $40'$, $31'$, 
$21'$ and $13'$, respectively (Bennett et al.\@ 2003). The resolution of WMAP at its lowest frequencies makes it 
unsuitable for our analysis of the W44 SNR.

In addition, it must be noted that a few examples of concave-down radio spectra of Galactic SNRs can be found in the 
recent literature, e.g.\@ cases of SNRs S147 and HB21. Xiao et al.\@ (2008) identified a spectral break at 1.5 GHz 
for SNR S147 and Pivato et al.\@ (2013) reported steepening at around 6 GHz in the case of SNR HB21. Finally, Planck 
Collaboration Int.\@ XXXI (2014) reported several other Galactic SNRs with possible high frequency spectral 
bending (such as IC443 and Puppis A). The high frequency spectral bending starts at relatively lower 
frequencies (several GHz) for S147 and HB21 than for SNRs W44, IC443 and Puppis A (several tens of GHz).

\centerline{\includegraphics[bb=0 0 504 360, width=1\columnwidth, keepaspectratio]{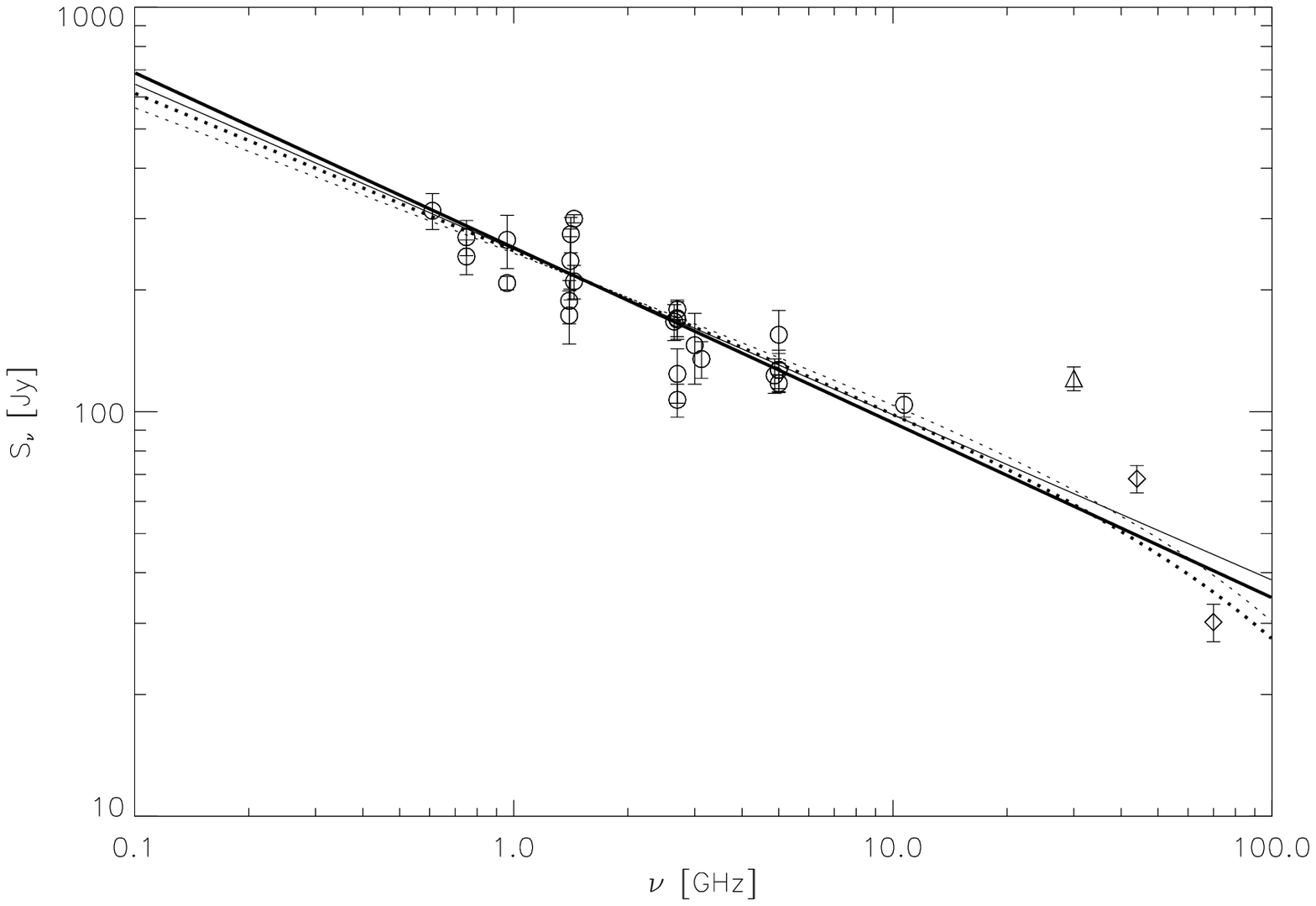}}
\figurecaption{1.}{The weighted least-squares fit to the data for the synchrotron power law model (solid line) as well as for 
synchrotron power law model with high frequency exponential cut-off (dotted line). Thick and thin lines correspond 
to the fits without and with inclusion of {\it Planck} data at 30 GHz, respectively. Diamond symbols indicate 
{\it Planck} data; the point at 30 GHz is shown as a triangle.}

\subsection{2.2. The analysis of possible spinning dust emission towards the SNR W44}

The spinning dust emission is currently one of the most probable proposed mechanisms to explain the so called anomalous 
microwave emission\footnote{Dust-correlated emission observed between around $10-100$ GHz that can not be accounted for 
by extrapolating the thermal dust emission to low frequencies.} (AME) from the diffuse interstellar medium (ISM) of 
Milky Way (Planck Collaboration Int.\@ XV 2014, and references therein). In fact, it is believed that the anomalous 
component of diffuse Galactic background is produced by electric dipole rotational emission from very small dust grains 
(Erickson 1957; Draine \& Lazarian 1998a). Small dust grains are likely to have a nonzero electric dipole moment due 
to the intrinsic dipole moment of molecules within the grain and uneven charge distribution. These grains will spin 
due to interaction with the ambient interstellar medium and radiation field, and thus radiate electromagnetic waves 
due to the rotation of their electric dipole moment (Ali-Ha\"{\i}moud et al.\@ 2009). To calculate the detailed 
frequency spectrum of spinning dust, one needs to integrate the emission over a distribution of grain sizes, electric 
dipole moments, and angular velocities (see Dickinson et al.\@ 2014, and references therein for more details). A lot 
of recent papers deal with additional improvements of the basic emission model (Ali-Ha\"{\i}moud et al.\@ 2009; 
Ysard et al.\@ 2010; Ysard \& Verstraete 2010; Hoang et al.\@ 2010, 2011; Ysard et al.\@ 2011).

Spinning dust emission can generally contribute significantly at high radio continuum frequencies, especially 
around 10 - 100 GHz (Draine \& Lazarian 1998ab; Ali-Ha\"{\i}moud et al.\@ 2009; Stevenson 2014; Planck Collaboration 
Int.\@ XV 2014). It is responsible for a characteristic bump in the high frequency part of the radio continuum.

Scaife et al.\@ (2007) fitted the radio spectrum of Galactic SNR 3C396 with a Warm Neutral Medium (WNM) 
spinning dust emission model (Draine \& Lazarian 1998b), asserting for the first time the possibility of 
spinning dust (rapidly rotating dust grains) emission from the vicinity of an SNR. Although the claims of possible 
contribution of thermal bremsstrahlung at high frequencies were addressed by Oni\'c et al.\@ (2012) for SNR 3C396, 
the same is not applicable in the case of the particular shape of radio spectrum of W44.

In their study of AME in Galactic clouds, Planck Collaboration Int.\@ XV (2014) listed W48 complex (that contains W44) as 
one of their candidate regions. However, their conclusions, based on a significance level of the AME detection, suggest 
that the excess emission, for this object, is not statistically highly significant. On the other hand, they emphasized 
that many of the so called statistically semi-significant AME regions (those at $2-5\sigma$), like W48, are likely 
to be real detections of AME. Planck Collaboration Int.\@ XV (2014) also noted that in general, the emerging picture is 
that the bulk of the AME (most probably due to spinning dust) is coming from the polycyclic aromatic hydrocarbons and 
small dust grains from the colder neutral interstellar medium phase. They also suggested that AME comes 
from the molecular cloud dust or PDR, but not from \mbox{H\,{\sc ii}} regions themselves. That is consistent 
with a general belief that PAHs are destroyed in \mbox{H\,{\sc ii}} regions, and the AME emissivity is 
lower in the ionized phase of the interstellar medium.

However, Hensley \& Draine (2015) voiced their concern on the validity of claims about AME/PAHs association. They proposed 
that one possibility is that AME could be in fact a spinning dust emission that arises primarily from very small grains 
that are not PAHs. Of course, AME could be a combination of the emission from spinning PAHs and non-PAH spinning dust, as 
well as the thermal dust emission, such as magnetic dipole emission (Hensley \& Draine 2015, and references therein). 
Hensley \& Draine (2015) did not find the evidence for strong AME correlation with free-free or CO emission, but they 
emphasize the plausibility of correlation with synchrotron emission. If the AME arises from spinning ultra small grains, 
it might be enhanced in SNRs in which grains are violently shattered. In spite of that, it can also be suppressed as the 
result of the destruction of very small grains by sputtering (Laki\'cevi\'c et al.\@ 2015).

In a recent paper, Irfan et al.\@ (2015) analyzed the region that encompasses SNR W44. The inner aperture size used for 
their aperture photometry was set to $60'$ and centered around the Galactic coordinates, $l=34\oo8$, $b=-0\oo5$. 
Irfan et al.\@ (2015) demonstrated the presence of AME associated with that region. They also ruled out the possibility 
that the excess emission near 30 GHz was from a nearby ultra-compact \mbox{H\,{\sc ii}} region.

As the proper physical conditions may exist, it is worth checking if spinning dust emission is significant enough 
to shape the radio spectrum of W44 SNR near 30 GHz. To that end, the SpDust code, ver.\@ 2.01 (Ali-Ha\"{\i}moud et al.\@ 2009; 
Silsbee et al.\@ 2011) was used. The radio continuum spectrum from 610 MHz to 70 GHz was fitted by the sum of synchrotron 
radiation represented by the power law with spectral index $\alpha$ and the spinning dust emission 
$S_\mathrm{spd}(\nu; n, T)$ (Eq.\@ 3)
\begin{equation}
S_{[\mathrm{Jy}]}(\nu) = S_{[\mathrm{Jy}]}^\mathrm{sync}(1\ \!\!\mathrm{GHz})\ \nu_{[\mathrm{ GHz}]}^{-\alpha} + S_\mathrm{spd}(\nu; n, T).                     
\end{equation}

A spinning dust emission spectrum from a WNM model (assume $n_\mathrm{H}=0.4\ \mathrm{cm}^{-3}$, $T=6000\ \mathrm{K}$) 
was used in fitting the overall radio continuum spectrum, as well as several other standard types of environments that were used for comparison: Warm Ionized Medium (WIM, 
$n_\mathrm{H}=0.1\ \mathrm{cm}^{-3}$, $T=8000\ \mathrm{K}$), Cold Neutral Medium (CNM, $n_\mathrm{H}=30\ \mathrm{cm}^{-3}$, 
$T=100\ \mathrm{K}$), and Molecular Cloud (MC, $n_\mathrm{H}=300\ \mathrm{cm}^{-3}$, $T=20\ \mathrm{K}$). These idealized ISM phases 
are defined in Table 1 of Draine \& Lazarian (1998b). The parameters for the grain size distribution are taken from Table 1 
of Weingartner \& Draine (2001) in accordance with the work of Ali-Ha\"{\i}moud et al.\@ (2009) - see Figure 14 of their 
work. Due to the low quality of radio spectrum we must confine ourselves to the qualitative analysis using standard ISM environments.

The weighted least-squares fit is calculated using the MPFIT\footnote{http://purl.com/net/mpfit} (Markwardt 2009) 
package written in IDL for all of the fits presented in this paper, with starting values estimated from the data. 
MPFIT also provides estimates of the 1$\sigma$ uncertainties for each parameter, taken as the square root of the diagonal 
elements of the parameter covariance matrix. We must note here that while modeling spinning dust emission can yield 
very good fits, one must exercise caution in interpreting these results since the fitted parameters can become unphysical.

Also, it is worth mentioning that in the case of non-linear model fitting, the number of degrees of 
freedom is generally unknown, i.e.\@ it is not possible to compute the value of reduced $\chi^{2}$, 
or adjusted $R^{2}$ (Andrae et al.\@ 2010). In non-linear models, the parameter $k =  N - p$ (where $N$ is 
the number of data points and $p$ is the number of model parameters) does not always represent the exact 
number of degrees of freedom (Andrae et al.\@ 2010).

}
\end{multicols}

\noindent
\parbox{\textwidth}{
\centerline{$
\begin{array}{cc}
\includegraphics[bb=0 0 504 360, width=0.5\columnwidth, keepaspectratio]{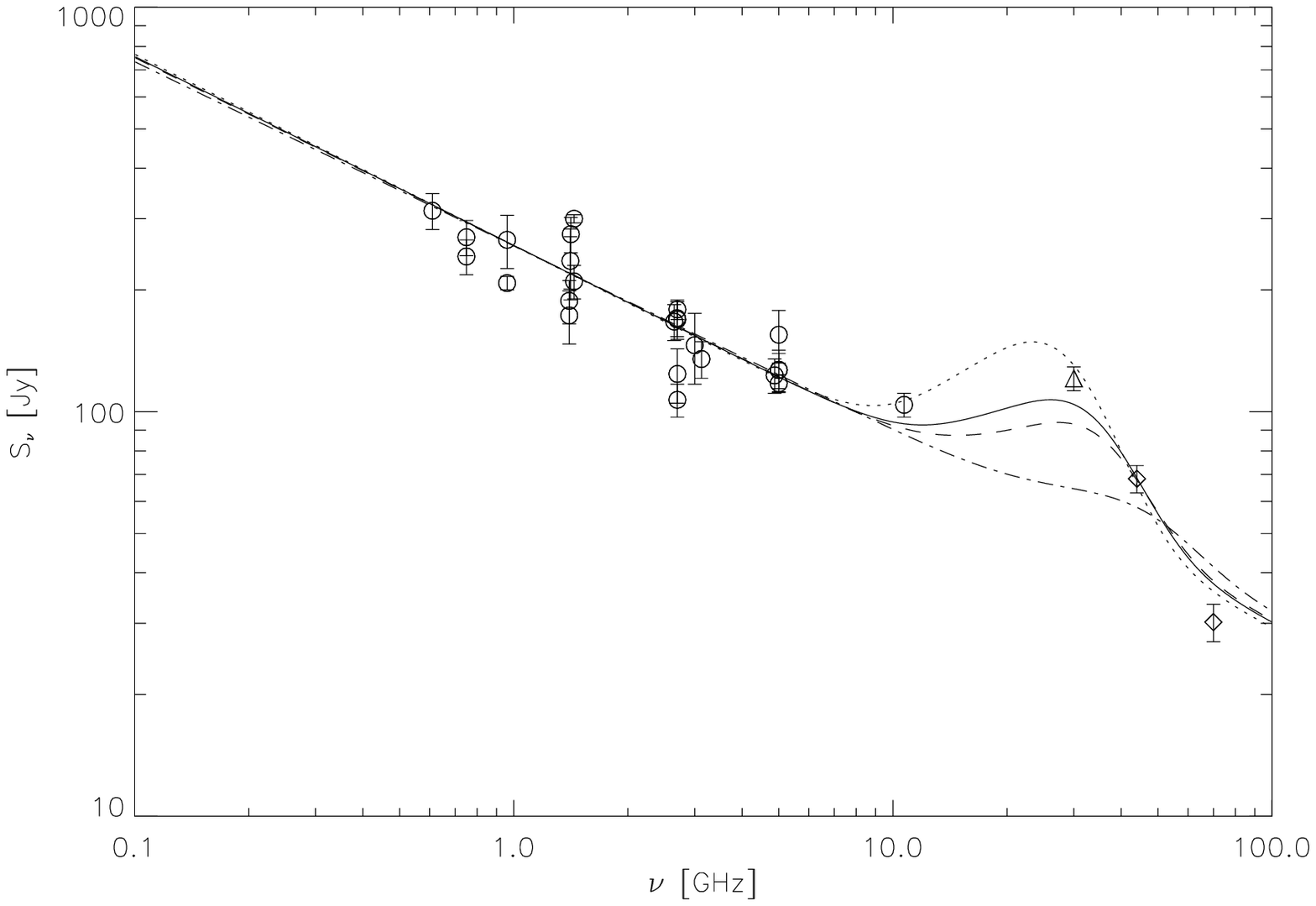} &
\includegraphics[bb=0 0 504 360, width=0.5\columnwidth, keepaspectratio]{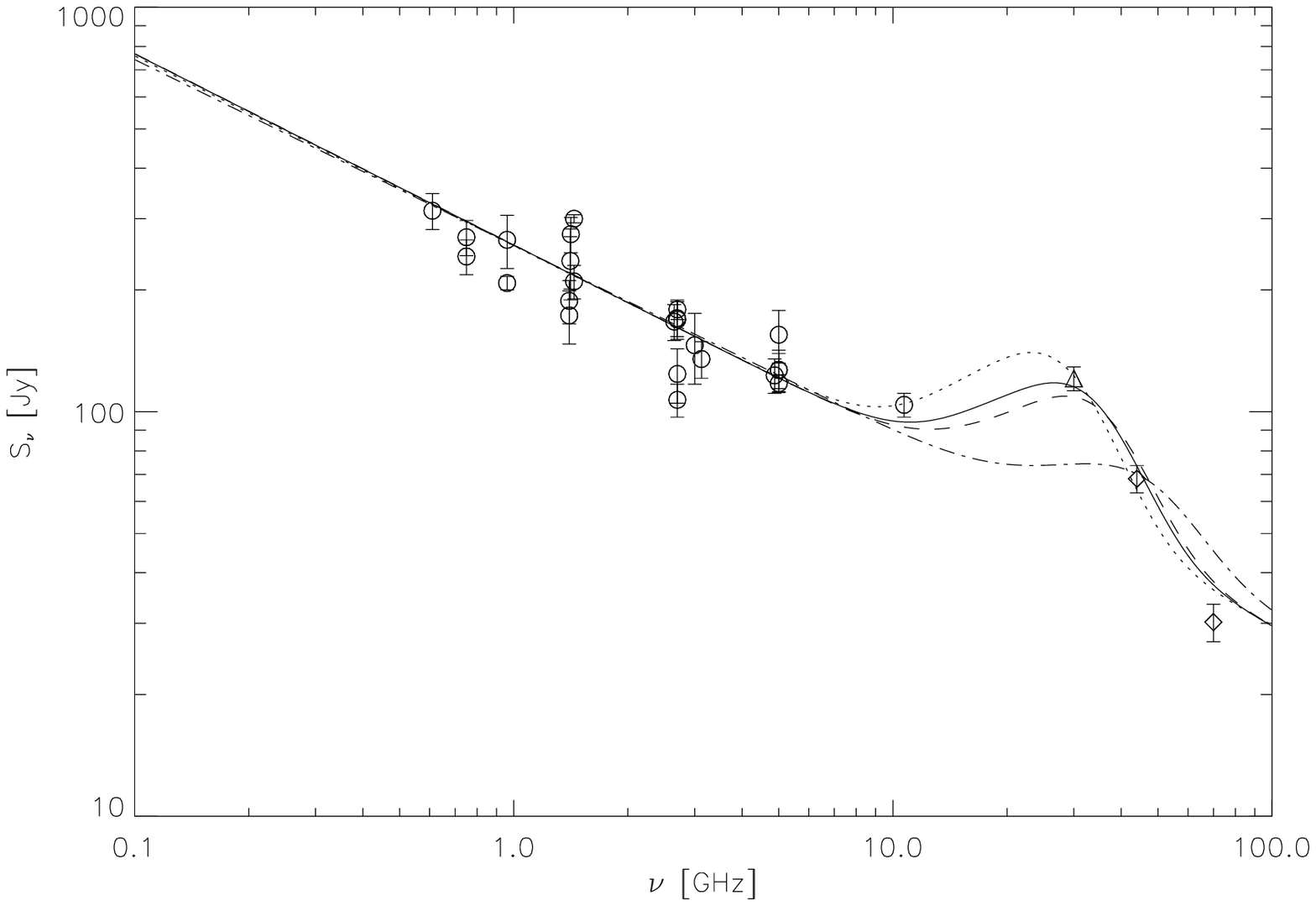} \\
\includegraphics[bb=0 0 504 360, width=0.5\columnwidth, keepaspectratio]{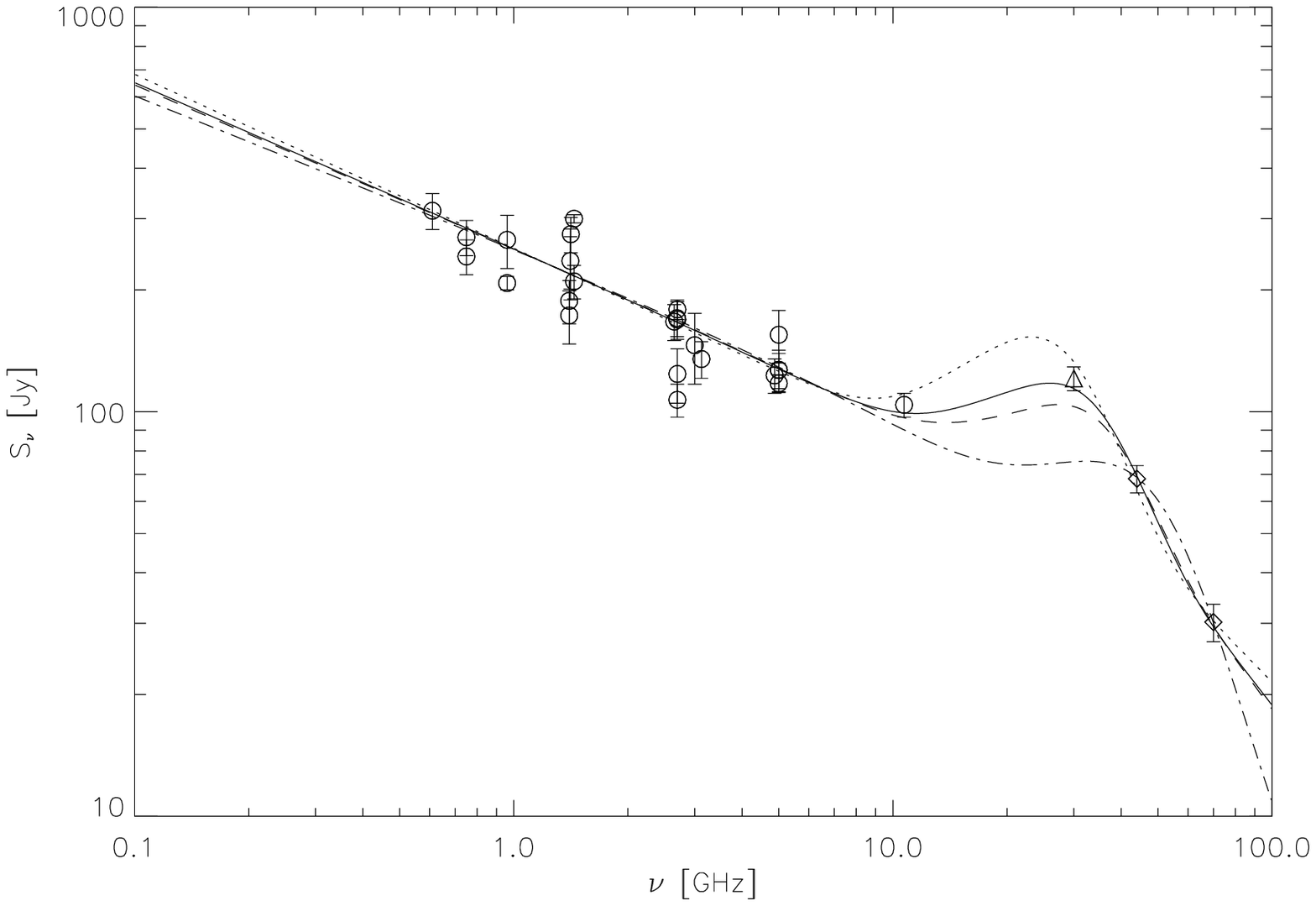} &
\includegraphics[bb=0 0 504 360, width=0.5\columnwidth, keepaspectratio]{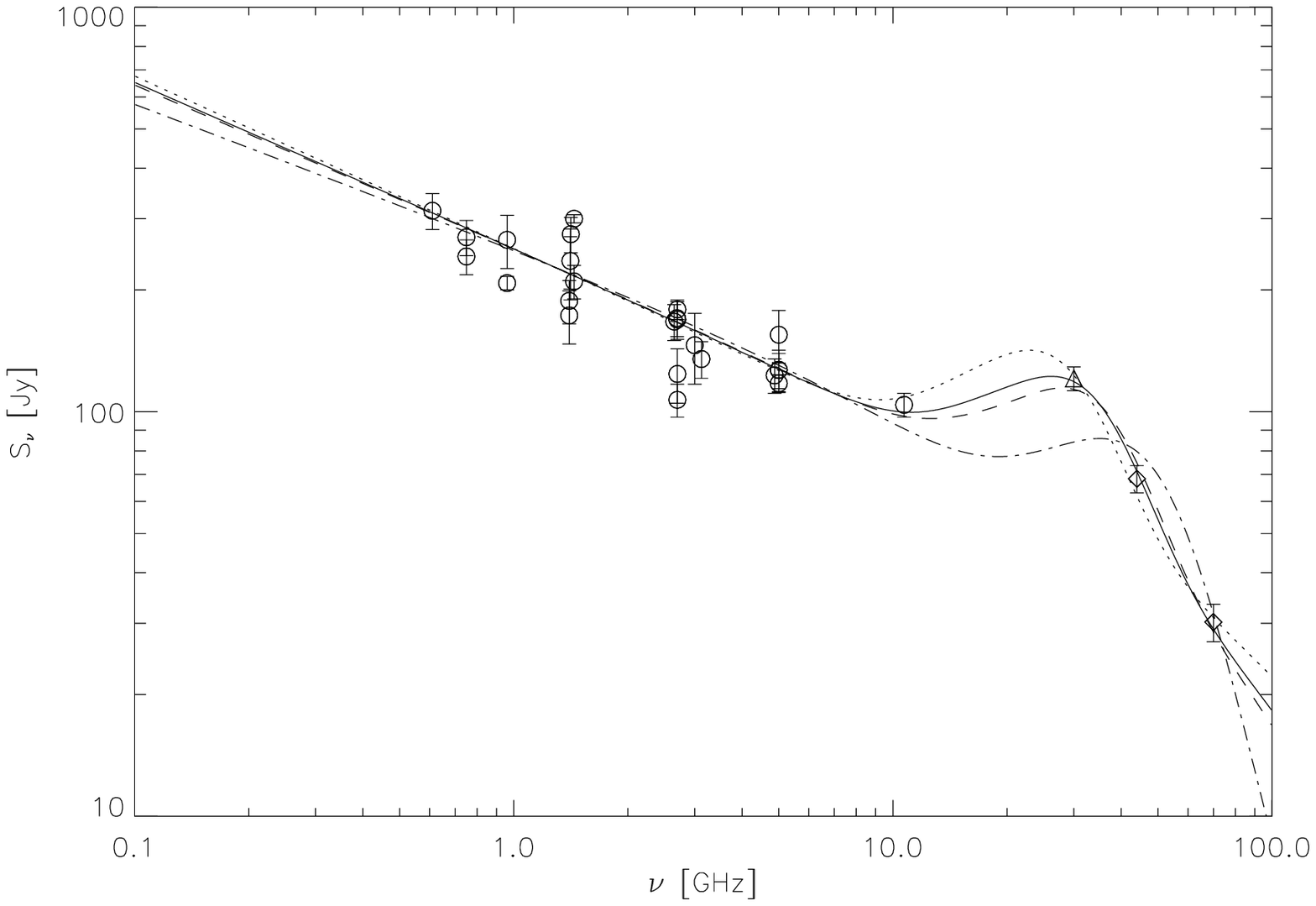}
\end{array}$}

\figurecaption{2.}{The weighted least-squares fit to the data for the synchrotron power law model with inclusion of spinning dust 
emission (Eq.\@ 3) is presented on the upper graphs and the weighted least-squares fit to the data for the synchrotron 
power law model with high frequency exponential cut-off with inclusion of spinning dust emission (Eq.\@ 4) is presented on 
the lower graphs. The left and right graphs correspond to the data samples with flux density at 30 GHz excluded and included, 
respectively. Diamond symbols indicate {\it Planck} data and triangle symbol represents the data point at 30 GHz. Different 
lines correspond to different models of idealized phases of interstellar medium: WIM - solid line, WNM - dotted line, 
CNM - dashed line, MC - dashed dot line.}}

\begin{multicols}{2}
{

The results of weighted least-squares fit to data by Eq.\@ 3 are presented in left and right upper graphs of Figure 2. 
The left and right graphs correspond to the data samples with flux density at 30 GHz excluded and included, respectively. 
Diamond symbols indicate {\it Planck} data and the triangle symbol represents data at 30 GHz. Different lines 
correspond to different models of idealized phases of interstellar medium: WIM - solid line, WNM - dotted line, CNM - dashed 
line, MC - dashed dot line. The best fitting parameters as well as corresponding $\chi^{2}\ (k)$ for different models are 
presented in Table 1 (star symbol indicates inclusion of the 30 GHz {\it Planck} data in the analysis).

The large scatter of flux densities measured at the same frequencies prevents us from making a firm discrimination 
between the models we are considering. We cannot rule out the significance of spinning dust emission from this SNR. 
A model that involves the WNM condition is statistically the most probable (supported by the standard analysis 
of $\chi^{2}/k$ values) and also physically plausible for the particular SNR environment. 
In addition, this model (dotted line in Figure 2) passes through the error bars of the data point 
at 30 GHz even when that point is excluded from the analysis. As it was noted by Planck Collaboration Int.\@ XXXI (2014), 
this point is possibly contaminated by unrelated large-scale emission from the Galactic plane. 
Although such a claim can not be fully dismissed, the strong possibility exists that the spinning dust emission 
is in fact, at least partially, responsible for the apparent bump around 30 GHz.

For completeness, the spinning dust flux density estimate $S_\mathrm{spd}(30)$, as well as its ratio to the total flux density 
$f(30)=S_\mathrm{spd}(30)/S(30)$ at 30 GHz is given in Table 1 for all considered models. The spinning dust fractions $f(30)$ 
are in accordance with the results obtained for other AME sources ($\approx0.5$) in Planck Collaboration Int.\@ XV (2014).

}\end{multicols}

\noindent
\parbox{\textwidth}{{\bf Table 1.} The best fitting parameters for different models.
\vskip.25cm \centerline{
\begin{tabular}{@{}lccccc@{}r}
\hline
Equation&$\alpha$&$\nu_{0}$ [GHz]&$S_\mathrm{ spd}(30)$ [Jy]&$f(30)$&$\chi^{2}\ (k)$\\
\hline
1&$0.433\pm0.014$&$-$&$-$&$-$&$268.18\ (26)$\\
1$^{*}$&$0.409\pm0.013$&$-$&$-$&$-$&$322.11\ (27)$\\
&&&\\
2&$0.388\pm0.024$&$234\pm109$&$-$&$-$&$262.84\ (25)$\\
2$^{*}$&$0.358\pm0.023$&$220\pm91$&$-$&$-$&$315.13\ (26)$\\
&&&\\
3 (WIM)&$0.467\pm0.018$&$-$&$51.6\pm12.9$&$0.50\pm0.20$&$251.76\ (25)$\\
3$^{*}$(WIM)&$0.473\pm0.017$&$-$&$63.5\pm7.7$&$0.55\pm0.12$&$253.13\ (26)$\\
3 (WNM)&$0.473\pm0.018$&$-$&$78.7\pm18.0$&$0.60\pm0.24$&$248.63\ (25)$\\
3$^{*}$(WNM)&$0.470\pm0.017$&$-$&$70.6\pm8.3$&$0.57\pm0.12$&$248.90\ (26)$\\
3 (CNM)&$0.465\pm0.018$&$-$&$40.1\pm10.6$&$0.43\pm0.18$&$253.60\ (25)$\\
3$^{*}$(CNM)&$0.473\pm0.017$&$-$&$57.1\pm7.2$&$0.52\pm0.12$&$258.53\ (26)$\\
3 (MC)&$0.456\pm0.019$&$-$&$9.7\pm5.0$&$0.15\pm0.10$&$264.09\ (25)$\\
3$^{*}$(MC)&$0.460\pm0.019$&$-$&$20.3\pm4.7$&$0.27\pm0.10$&$303.06\ (26)$\\
&&&\\
4 (WIM)&$0.408\pm0.025$&$136\pm46$&$63.2\pm13.7$&$0.55\pm0.23$&$241.14\ (24)$\\
4$^{*}$(WIM)&$0.409\pm0.025$&$130\pm42$&$68.4\pm8.0$&$0.58\pm0.15$&$241.35\ (25)$\\
4 (WNM)&$0.428\pm0.027$&$196\pm88$&$81.7\pm18.3$&$0.62\pm0.26$&$242.93\ (24)$\\
4$^{*}$(WNM)&$0.424\pm0.026$&$203\pm91$&$71.3\pm8.4$&$0.58\pm0.15$&$243.34\ (25)$\\
4 (CNM)&$0.403\pm0.025$&$127\pm42$&$52.0\pm11.5$&$0.51\pm0.21$&$241.83\ (24)$\\
4$^{*}$(CNM)&$0.402\pm0.025$&$112\pm33$&$64.0\pm7.7$&$0.56\pm0.15$&$243.75\ (25)$\\
4 (MC)&$0.374\pm0.028$&$61\pm22$&$31.5\pm7.9$&$0.42\pm0.22$&$243.32\ (24)$\\
4$^{*}$(MC)&$0.354\pm0.031$&$47\pm17$&$43.4\pm8.1$&$0.52\pm0.23$&$268.05\ (25)$\\
\hline
\end{tabular}}}

\bigskip

\centerline{$^{*}$For the case when {\it Planck's} data point at 30 GHz is included in the analysis.}

\vskip.5cm

\begin{multicols}{2}
{

Additional support for a spinning dust emission hypothesis can be found in its correlation with thermal dust 
emission, especially at IRAS (Infrared Astronomical Satellite) wavelengths. Planck Collaboration Int.\@ XV (2014) found that AME 
sources generally have a $12\ \mu\mathrm{m}$/$25\ \mu\mathrm{m}$ ratio $\approx(0.6-1.0)$. This is interpreted as a confirmation 
of the spinning dust model, where the very small grains are responsible for the bulk of the AME. Although the W44 SNR 
was not detected by Saken et al.\@ (1992), Arendt (1989) found a $12\ \mu\mathrm{m}$/$25\ \mu\mathrm{ m}$ 
ratio of $\approx1$. However, possibility of ionic line contamination of the particular IRAS bands can not be fully dismissed 
(Oliva et al.\@ 1999; Reach \& Rho 1996).

The comparison was also made between the spinning dust flux density estimates at 30 GHz $S_\mathrm{spd}(30)$ for different models to the 
$100\ \mu\mathrm{m}$ (3000 GHz) flux density $S(3000)$ as given in Arendt (1989). The ratio $S_\mathrm{spd}(30)/S(3000)$ 
spans a large range of values for the different environmental models: $(7-60)\times10^{-4}$. Bearing in mind the roughness 
of our analysis, as well as high uncertainty of the IRAS $100\ \mu\mathrm{ m}$ flux density (around 50\%), our values 
for $S_\mathrm{spd}(30)/S(3000)$ are in a rough agreement (slightly higher) with other determinations for AME sources 
($(1-15)\times10^{-4}$, Planck Collaboration Int.\@ XV 2014; Hensley \& Draine 2015, and references therein).

Finally, it is tempting to check whether the fits will improve if the model that incorporates the sum of power law with 
exponential cut-off and spinning dust emission is used (Eq.\@ 4)\begin{equation}
S_{[\mathrm{Jy}]}(\nu) = S_{[\mathrm{Jy}]}^\mathrm{sync}(1\ \!\!\mathrm{GHz})\ \nu_{[\mathrm{GHz}]}^{-\alpha}\ e^{-\frac{\nu}{\nu_{0}}} + S_\mathrm{spd}(\nu; n, T).     
\end{equation}

The lower graphs of Figure 2 represent the weighted least-squares fit to the data for the synchrotron power law 
model with high frequency exponential cut-off with inclusion of spinning dust emission (Eq.\@ 4). The left and right 
graphs correspond to the data samples with flux density at 30 GHz excluded and included, respectively. The best fitting 
parameters as well as corresponding $\chi^{2}\ (k)$ for different models are given in Table 1.

The average $\chi^{2}/k$ is slightly less for Eq.\@ 4 than for Eq.\@  3 (see Table 1). On the other hand, statistical 
discrimination between different environmental conditions used for Eq.\@ 4 is in this case even more inconclusive. Still, 
detection of hard X-ray emission from W44, not connected with associated pulsar or its nebula, reported by 
Uchida et al.\@ (2012), can cast doubt that high frequency synchrotron cut-off at around 100 - 200 GHz (see Table 1) is 
genuine and is primarily responsible for a lower flux density at 70 GHz. On the other hand, Lee et al.\@ (2015) recently explored 
the non-thermal emission mechanisms of dynamically evolved SNRs. They explored two scenarios of particle acceleration, either 
a re-acceleration of Galactic cosmic rays, or an efficient nonlinear diffusive shock acceleration of thermally injected 
particles by a fast radiative cloud shock. Lee et al.\@ (2015) emphasized that if sufficiently strong magnetic turbulence 
is present in the molecular cloud, the re-acceleration scenario of the non-thermal emission agrees well with the broadband 
spectrum of SNR W44 and that the apparent discrepancy at 30 GHz can possibly arise from the anomalous microwave emission 
from small spinning dust grains. Their predicted radio spectrum shows a spectral softening above around 10 GHz due to 
synchrotron loss, and is consistent with the 70 GHz flux density from {\it Planck}.

One must bear in mind that SNR W44 is actually placed in a very complex region (such that physical conditions 
inside the remnant may significantly vary) so that the more detailed models than these presented in this paper 
should be more adequate. In fact, thorough radiative transfer modeling is required to further our understanding 
of the spinning dust emission and its significance. Finally, statistically insufficient number of data prevents 
us to make a firm quantitative discrimination between different models. In addition to the general need for more 
data, the much more improved resolution ($<1'$) of new observations at radio continuum frequencies between 
10 and 100 GHz would also be beneficial.

\section{3. CONCLUSIONS}

In this paper, the integrated radio spectrum of SNR W44 was analyzed up to 70 GHz testing the different emission 
models that can be responsible for its particular shape. The main conclusions are:

\bigskip

\item{(1)} {\it Planck} observations made it possible to analyze the high frequency part of radio emission from SNRs.

\item{(2)} Although the quality of radio continuum spectrum (a high scatter of data points at same frequencies) prevents 
us to make definite conclusions, the possibility of spinning dust emission detection towards this remnant is emphasized. 
In fact, the spinning dust emission is proposed to be, at least partially, responsible for the apparent bump in the 
radio continuum around 30 GHz.

\item{(3)} In addition, a concave-down feature in the radio spectrum, due to synchrotron losses, can not be definitely 
dismissed by the present knowledge of the radio continuum spectrum of this SNR. 

\item{(4)} Finally, the more data at different frequencies between 10 and 100 GHz at much better resolution than 
{\it Planck} are needed to make firm conclusions about the contribution of particular radiation mechanisms responsible 
for an observed shape of the radio spectrum of SNR W44.


\acknowledgements{I want to thank Dejan Uro\v sevi\'c for careful reading of the manuscript and useful comments that 
substantially improved this paper. I am also grateful to T.\@ Hoang as well as to the anonymous referee for valuable 
suggestions. This work is part of Project No. 176005 "Emission Nebulae: Structure and Evolution" supported by the 
Ministry of Education, Science, and Technological Development of the Republic of Serbia.}


\references

Ali-Ha\"{\i}moud, Y., Hirata, C. and Dickinson, C.: 2009, \journal{Mon. Not. R. Astron. Soc.}, \vol{395}, 1055.

Altenhoff, W. J., Downes, D., Goad, L., Maxwell, A. and Rinehart, R.: 1970, \journal{Astron. Astrophys}, \vol{1}, 319.

Anderl, S., Gusdorf, A. and G\"{u}sten, R.: 2014, \journal{Astron. Astrophys}, \vol{569}, 81.

Andrae, R., Schulze-Hartung, T. and Melchior, P.: 2010, arXiv:1012.3754.

Arendt, R. G.: 1989, \journal{Astrohys. J. Suppl. S.}, \vol{70}, 181.

Baars, J. W. M, Genzel, R., Pauliny-Toth, I. I. K. and Witzel, A.: 1977, \journal{Astron. Astrophys}, \vol{61}, 99.

Bennett, C. L. et al.: 2003, \journal{Astrophys. J.}, \vol{583}, 1.

Cardillo, M., Tavani, M., Giuliani, A., Yoshiike, S., Sano, H., Fukuda, T., Fukui, Y., Castelletti, G. and 
Dubner, G.: 2014, \journal{Astron. Astrophys}, \vol{565}, 3754.

Castelletti, G., Dubner, G., Brogan, C. and Kassim, N.E.: 2007, \journal{Astron. Astrophys}, \vol{471}, 537.

Dickinson, C. Ali-Ha\"{\i}moud, Y., Beswick, R. J., Casassus, S., Cleary, K., Draine, B. T., Genova-Santos, R., 
Grainge, K., Hoang, T. C., Lazarian, A. et al.: 2014, arXiv:1412.5054.

Draine, B. T. and Lazarian A.: 1998a, \journal{Astrophys. J. Lett.}, \vol{494}, L19.

Draine, B. T. and Lazarian A.: 1998b, \journal{Astrophys. J.}, \vol{508}, 157.

Erickson, W. C.: 1957, \journal{Astrophys. J.}, \vol{126}, 480.

Green D. A., 2014, 'A Catalogue of Galactic Supernova Remnants (2014 May version)', 
Cavendish Laboratory, Cambridge, United Kingdom (available at\\ 
"http://www.mrao.cam.ac.uk/surveys/snrs/").

Hensley, B. S. and Draine, B. T.: 2015, arXiv:1505.02157.

Hoang, T., Draine, B. T. and Lazarian, A.: 2010, \journal{Astrophys. J.}, \vol{715}, 1462.

Hoang, T., Lazarian, A. and Draine, B. T.: 2011, \journal{Astrophys. J.}, \vol{741}, 87.

Irfan, M. O., Dickinson, C., Davies, R. D. et al.: 2015, \journal{Mon. Not. R. Astron. Soc.}, \vol{448}, 3572.

Laki\'cevi\'c, M., van Loon, J. T., Meixner, M. et al.: 2015, \journal{Astrophys. J.}, \vol{799}, 50.

Lee, S-H., Patnaude, D. J., Raymond, J. C., Nagataki, S., Slane, P. O. and Ellison, D. C.: 2015, \journal{Astrophys. J.}, 
\vol{806}, 71

Mandolesi, N., Bersanelli, M., Butler, R. C. et al.: 2010, \journal{Astron. Astrophys}, \vol{520}, 3.

Markwardt, C. B.: 2009, ASP. Conf. Ser., \vol{411}, 251.

Oliva, E., Lutz, D., Drapatz, S. and Moorwood, A. F. M., 1999, \journal{Astron. Astrophys}, \vol{341}, 75.

Oni\'c, D.: 2013, \journal{Astrophys. Space Sci.}, \vol{346}, 3.

Oni\'c, D., Uro\v sevi\'c, D., Arbutina, B. and Leahy, D.: 2012, \journal{Astrophys. J.}, \vol{756}, 61.

Ortega, M. E., Paron, S., Cichowolski, S., Rubio, M., Castelletti, G. and Dubner, G.: 2010, \journal{Astron. Astrophys}, \vol{510}, 96.

Paron, S., Ortega, M. E., Rubio, M. and Dubner, G.: 2009, \journal{Astron. Astrophys}, \vol{498}, 445.

Pivato, G., Hewitt, J. W., Tibaldo, L., Acero, F., Ballet, J., Brandt, T. J., de Palma, F., Giordano, F., Janssen, 
G. H., J\'{o}hannesson, G. and Smith, D. A., 2013, \journal{Astrophys. J.}, \vol{779}, 179.

Planck Collaboration Int.\@ XV: Ade, P. A. R., Aghanim, N., Alves, M. I. R. et al.: 2014, \journal{Astron. Astrophys}, \vol{565}, 103.

Planck Collaboration Int.\@ XXXI: Arnaud, M., Ashdown, M., Atrio-Barandela, F. et al.: 2014, arXiv:1409.5746.

Reach, W. T. and Rho, J.: 1996, \journal{Astron. Astrophys}, \vol{315}, 277.

Rho, J. and Petre, R.: 1998, \journal{Astrophys. J.}, \vol{503}, L167.

Saken, J. M., Fesen, R. A. and Shull, J. M.: 1992, \journal{Astrohys. J. Suppl. S.}, \vol{81}, 715.

Scaife, A., Green, D. A., Battye, R. A. et al.: 2007, \journal{Mon. Not. R. Astron. Soc.}, \vol{377}, L69.

Silsbee, K., Ali-Ha\"{\i}moud, Y. and Hirata C.: 2011, \journal{Mon. Not. R. Astron. Soc.}, \vol{411}, 2750.

Stevenson, M. A.: 2014, \journal{Astrophys. J.}, \vol{781}, 113.

Sun, X. H., Reich, P., Reich, W., Xiao, L., Gao, X. Y. and Han, J. L.: 2011, \journal{Astron. Astrophys}, \vol{536}, 83.

Uchida, H., Koyama, K., Yamaguchi, H., Sawada, M., Ohnishi, T., Tsuru, T., Go., Tanaka, T., Yoshiike, S. and Fukui, Y.: 2012, 
\journal{Publ. Astron. Soc Jpn.}, \vol{64}, 141.

Uro\v sevi\'c, D.: 2014, \journal{Astrophys. Space Sci.}, \vol{354}, 541.

Vink, J.: 2012, \journal{Astron. Astrophys. Rev.}, \vol{20}, 49.

Weingartner J. C. and Draine B. T.: 2001, \journal{Astrophys. J.}, \vol{548}, 296.

Xiao, L., F\"{u}rst, E., Reich, W. and Han, J. L., 2008, \journal{Astron. Astrophys}, \vol{482}, 783.

Ysard, N., Miville-Desch\^{e}nes, M. A. and Verstraete, L.: 2010, \journal{Astron. Astrophys}, \vol{509}, 1.

Ysard, N. and Verstraete, L.: 2010, \journal{Astron. Astrophys}, \vol{509}, 12.

Ysard, N., Juvela, M. and Verstraete, L.: 2011, \journal{Astron. Astrophys}, \vol{535}, 89.

Yoshiike, S., Fukuda, T., Sano, H., Ohama, A., Moribe, N., Torii, K., Hayakawa, T., Okuda, T., Yamamoto, H., 
Tajima, H. et al.: 2013, \journal{Astrophys. J.}, \vol{768}, 179.

\endreferences

}
\end{multicols}

\vfill\eject

 {\ }
 
 
 
\naslov{O INTEGRALNOM NEPREKIDNOM RADIO-SPEKTRU OSTATKA SUPERNOVE {\rm \bf W44 (G}34.7-0.4): NOVA SAZNA{\NJ}A 
POMO{\CJ}U PLANKA}
 
 
\authors{D. Oni{\' c}}
\vskip3mm
 
 
\address{Department of Astronomy, Faculty of Mathematics,
University of Belgrade\break Studentski trg 16, 11000 Belgrade, Serbia}

\Email{donic}{math.rs}
 
\vskip.7cm
 
%
%
%
 
\vskip.7cm
 
\begin{multicols}{2}
{

{\rrm U ovom radu je analiziran integralni neprekidni radio-spektar ostatka supernove (OSN) 
{\rm W}44 sve do 70 {\rm GHz} i testirani su razli{\ch}iti emisioni modeli koji mogu
uzrokovati njegov svojevrsni oblik. Posmatranja realizovana svemirskim teleskopom Plank omogu{\cj}ila 
su da se analizira visokofrekventni deo radio-emisije OSN. Iako je kvalitet neprekidnog
radio-spektra (veliko rasipanje vrednosti na istim frekvencijama) takav da nam
onemogu{\cj}ava da izvedemo {\ch}vrste zaklju{\ch}ke, istaknuta je najverovatnija detekcija emisije
rotiraju{\cj}e pra{\sh}ine u pravcu ovog ostatka. Uz to, trenutno poznavanje integralnog
neprekidnog radio-spektra ovog OSN nije dovoljno da bi se definitivno mogla
osporiti tvrdnja o konkavnom krivljenju spektra usled sinhrotronskih gubitaka.}
 
}
\end{multicols}

\end{document}